\documentstyle [12pt]{article}

\newlength{\extraspace}
\newlength{\extraspaces}
\newcommand{\beq}{\begin{equation}
\addtolength{\abovedisplayskip}{\extraspaces}
\addtolength{\belowdisplayskip}{\extraspaces}
\addtolength{\abovedisplayshortskip}{\extraspace}
\addtolength{\belowdisplayshortskip}{\extraspace}}
\newcommand{\eeq}{\end{equation}}

\newcommand{\beqa}{\begin{eqnarray}
\addtolength{\abovedisplayskip}{\extraspaces}
\addtolength{\belowdisplayskip}{\extraspaces}
\addtolength{\abovedisplayshortskip}{\extraspace}
\addtolength{\belowdisplayshortskip}{\extraspace}}
\newcommand{\eeqa}{\end{eqnarray}}

\begin{document}

\begin{flushright}
{\sc NUS/HEP/}920501\\
May 1992
\end{flushright}
\vspace{.3cm}

\begin{center}
{\Large{\bf{\protect{$c=1$} Conformal Field Theory and\\[2mm]
the Fractional Quantum Hall Effect}}}\\[13mm]

{\sc Christopher Ting}\footnote{e-mail: scip1005@nuscc.nus.sg}\\[2mm]
{\it Defence Science Organization,
20 Science Park Drive, Singapore 0511} \\[2mm]
{and}\\[2mm]
{\it Department of Physics,
National University of Singapore\\[2mm]
Lower Kent Ridge Road, Singapore 0511} \\[2mm]
{and}\\[2mm]
{\sc C. H. Lai}\footnote{e-mail: phylaich@nusvm.bitnet}\\[2mm]
{\it Department of Physics,
National University of Singapore\\[2mm]
Lower Kent Ridge Road, Singapore 0511}
\\[10mm]
{\bf Abstract}
\end{center}
\noindent
We examine the application of $c=1$ conformal field theory to
the description of the fractional quantum Hall effect (FQHE).
It is found that the
Gaussian model together with an appropriate boundary condition
for the order parameter furnishes an effective theory for the
Laughlin type FQHE. The plateau formation condition corresponds to
taking the {\em chiral} portion of the theory.\par
\vfil

\newpage
Two-dimensional conformal field theories describe statistical systems at
criticality and furnish the classical solutions of string theory.
Recently, it has been suggested
that the order parameter of the fractional quantum
Hall effect (FQHE) is related to the vertex operator, and the
ground state wavefunction of a certain fractional filling factor
can be expressed in terms of the $N$-point
correlation function of vertex operators \cite{Fubini,MR}.
The application of conformal field theory has thus been extended into
a rather peculiar condensed matter phenomenon.

While the notion of criticality in the fractional quantum Hall effect is
still quite ambiguous, the vertex operator
approach to constructing wavefunctions provides a convenient
platform to discuss the statistics of the quasiparticle-quasihole
excitation. The physical picture behind the FQHE in the conformal field theory
language is that the locations at which vertex operators are inserted
correspond to the attachment of flux tubes. Since these locations are where
the electrons reside, the flux tubes can move around when electrons
traverse the 2-dimensional interface or heterojunction.

{}From a related perspective, we have
considered the electrons in the ``puncture phase" wherein each
electron sees the rest as punctures and therefore the configuration space
available to each is no longer ${\bf R}^2$ \cite{Wu,Lai-Ting}.
Because the configuration space has become
non-simply connected, the relevant permutation group for two particles
switching positions generalizes to the braid group in 2 dimensions.
This is the theoretical basis for the existence of anyons \cite{Wilczek}.

The link between anyon and conformal field theory is
revealed in the representation theory of the braid group.
It has been shown that the topological effect of the
``puncture phase'' is to introduce a delta-function potential
into the Hamiltonian \cite{Ting-Lai-1,Ting-Lai-2}.
In other words, except for a finite number of points,
the electrons are still non-interacting.
{}From the Hamiltonian describing the braid group statistics,
it turns out that the Knizhnik-Zamolodchikov equations \cite{KZ}
are the ground state equations for the particles in the ``puncture phase''.
Solving these equations with a set of appropriate boundary conditions
amounts to the fact that the conformal blocks of the 2-dimensional
conformal field theory
can be taken as the wavefunctions
of the strongly correlated many-body quantum mechanical system.

The physical considerations of the FQHE
entail that the ground state wavefunctions $\psi$
be analytic functions up to a
damping factor. In other words,
\beq
\psi \equiv f ( z_1, \cdots , z_N )
\exp ( - \frac{1}{4 \ell^2} \sum_i^N |z_i|^2 ) \, ,
\label{gs}
\eeq
and
$f ( z_1, \cdots , z_N )$ is a complex analytic function.
$\ell = \sqrt{ \frac{\hbar}{e B} }$ denotes
the magnetic length which is the radius of the cyclotron motion of
a particle with electric charge $e$ under the influence of an
uniform background magnetic field of strength $B$.
Having a holomorphic $f$ means that the mixing from other
Landau levels is negligible, as the
energy gap between the levels is large.
Now, it occurs that $f ( z_1, \cdots, z_N)$ is given by some
conformal block of a conformal field theory
\cite{Fubini} with $c=1$:
\beq
f ( z_1, \cdots , z_N ) =
\langle p_{total} | \prod_j^N e^{i p \phi (z_j)} | 0 \rangle \, ,
\eeq
where $p_{total} = N p$ is the charge neutrality condition.
The ``out'' state
\beq
\langle p_{total} | \equiv \lim_{z \rightarrow \infty} z^{p^2_{total}} \,
\langle 0 | \, e^{- i p_{total} \phi (z) }
\eeq
can be interpreted as a state characterized by the charges of a classical
plasma at the edge of the sample.
In the braid group approach,
$\psi$
is the {\em exact} solution to the ground state equation in the Schr\"odinger
picture \cite{Ting-Lai-1}:
\beq
\left( \partial_{z_j}
+ \frac{e B}{4 \hbar c} {\overline{z}_{j}}
- q \sum_{k=1, k \neq j}^N \frac{1}{z_j - z_k} \right)
\psi= 0
\eeq
for every $j$.
When we write $\psi$ as (\ref{gs}), the above equation is reduced to
the Knizhnik-Zamolodchikov equation for which $f(z_1, \cdots, z_N)$
is the solution. Since for $U(1)$ charges, the space of conformal blocks
is determined by the
Knizhnik-Zamolodchikov equation, we have thus established that the
holomorphic part of the wavefunction $\psi$, namely
$f(z_1, \cdots, z_N)$, can be identified as the conformal block of
the $U(1)$ current algebra.
We believe this is how the conformal blocks become relevant in
the FQHE \cite{Ting-Lai-2}.
The exponential factor of $\psi$ is essential here
as it enables the ground state to be a legitimate {\em normalizable}
wavefunction.
Notice that this factor is not directly accounted for in the string
theory approach \cite{Fubini}. It is put in
place by assuming that the measure is defined by
\beq
d \mu = \prod_i^{N} \exp ( - \frac{1}{4 \ell^2} |z_i|^2 ) \, d^2 z_i \, .
\label{measure}
\eeq
The requirement for normalizability highlights the fact that
the magnetic field in the background plays a crucial part in FQHE, which
has been emphasized in \cite{Ting-Lai-2}.

Having said that, it is worthwhile to note that
there is something extraordinary about
$f(z_1, \cdots, z_N)$ that distinguishes the FQHE
from the usual Hall effect with
integral filling factors. In the case of the FQHE with filling factor
$\frac{1}{q}$ where $q$ is an odd number,
\beq
f(z_1, \cdots, z_N) = {\rm const.} \, \prod_{ i < j }^N ( z_i - z_j )^q \, .
\eeq
That this Laughlin ans\"atz \cite{Laughlin}
satisfies the Knizhnik-Zamolodchikov equation
\beq
\left( \partial_{z_j}
- q \sum_{k=1, k \neq j}^N \frac{1}{z_j - z_k} \right)
f(z_1, \cdots, z_N, ) = 0
\eeq
is readily verifiable. The 2-dimensional conformal field
theory relevant to this class of FQHE ground states is the $c=1$ theory.

Now the field-theoretic representation of
the $U(1)$ current algebra is just the Gaussian model,
a boson $\varphi (z, {\overline z})$ compactified on
a circle of (arbitrary) radius $R$, and it has $c=1$.
The previous arguments allow one to take the action of the Gaussian model
\beq
S [ \varphi ]_R = \frac{\lambda}{2 \pi} \int d^2 z \,
\partial \varphi \overline{\partial}
\varphi
\label{Gaussian}
\eeq
as the {\em effective} theory describing the ground states of the FQHE.
$\lambda$ is the dimensionless constant and it will be normalized to 1
subsequently in the calculation of correlation function.
The salient feature of a $c=1$ theory is that there exists an {\em integrable}
marginal operator that changes $R$ in a continuous fashion.
The moduli space of the
$c=1$ theories corresponding to different values of $R$
is indexed by two critical lines called the Gaussian line and the
orbifold line respectively. The Gaussian line is generated by the
marginal operator perturbing the Gaussian model while the orbifold line
is associated with theories with the additional symmetry
$\varphi \rightarrow - \varphi$.
{}From the point of view of statistical mechanics, the radius $R$ corresponds
to the coupling constant of the statistical model.
Changing $R$ is equivalent to
changing the critical exponents of the order parameters, and hence
leads to different statistical models at criticality
\cite{Kadanoff,Ginsparg,DVV,Ginsparg-1}.

To elucidate this point, consider the vertex operators of the
$U(1)$ theory:
\beq
V_{n m} = \exp ( i p \phi + i {\overline p} {\overline \phi} ) \, ,
\eeq
where
\beq
\varphi (z, {\overline z}) = \phi (z) + \overline \phi (\overline z ) \, .
\eeq
The $U(1)$ charge ( or the momentum in the context of string theory )
lives on an even and self-dual lattice:
\beq
\Gamma_R = \left\{ ( p, {\overline p} ) = \left( \frac{m}{2R} + nR, \,
\frac{m}{2R} - nR \right); \, n, m \in Z \right\} \, .
\label{lattice}
\eeq
The $c=1$ theory is invariant with respect to a duality
transformation $ R \leftrightarrow \frac{2}{R}$,
$ m \leftrightarrow n $, as
can be seen from (\ref{lattice}) directly.
The conformal weight of $V_{nm}$ is $( h_{nm}, \overline{h}_{nm} )
= ( \frac{p^2}{2}, \frac{\overline{p}^2}{2} )$.
The assumption of no mixing from other Landau levels is encoded
in this language as taking the {\em chiral} half of the spectrum,
so that indeed $f$ is holomorphic.
In other words, $\overline p$ vanishes.

Thus, the vanishing of $\overline p$ seems to correspond to the fact that
{\em all} the electrons are in the lowest Landau level.
In this letter, we want to provide yet another insight on
why ${\overline p} = 0$ is the right way to describe the FQHE.

The defining characteristic of the FQHE is none other
than the appearance of plateaux at fractional
filling factors. Collectively, the electrons acquire the extra capacity
to buffer the surplus and deficit in the filling factor over a finite range
of magnetic field strength. According to Laughlin's theory \cite{Laughlin},
quasiparticle-quasihole excitations are admissble and
their creations require a finite amount of energy. Precisely because these
excitations are not gapless, plateaux become observable.
The question we would like to pose is the following:
Can one describe the formation of a plateau
in the language of conformal field theory?
The key to address this question lies in the fact that we are dealing with
a $c=1$ theory, for which an {\em integrable} marginal operator exists.

The quantum mechanics of a system of 2-dimensional electrons in an external
magnetic field applied perpendicular to the plane tells us that each
Landau level is degenerate, namely it can accommodate
$\frac{e B}{ 2 \pi \hbar}$ per unit area of electrons. In a typical
experimental setup, the number density $\rho$ is fixed and the filling
factor $\nu = \frac{ 2 \pi \hbar \rho}{e B}$.
Therefore $q = \frac{1}{\nu}$
is directly proportional to the magnetic field strength $B$.

Now in Laughlin's picture, the spins of the electrons are frozen
in the direction of the magnetic field and the field strength being
strong, the flipping fluctuation can be taken to be absent.
Given that the Gaussian model is the effective theory for the
ensemble of electrons in the FQHE phase,
the ``coupling constant'' $\lambda$ can then be seen as
the magnetic field strength $B$ that drives the
original many-body quantum mechanical system.
The correspondence is justified on the ground that the filling
factor depends only on $B$ while the only parameter of the
Gaussian model is $\lambda$. Clearly, the effective action (which is the
effective Hamiltonian of the underlying microscopic system in the
continuum limit) is not valid for all $B$. It is postulated to describe
the FQHE. Thus, we suppose that the center of a particular plateau is at
$B_0$, and the width is $\delta B$. Translated in the language of the
effective theory, we are only concerned with a reference coupling constant
$\lambda_0$ corresponding to $B_0$ and the {\em finite} interval
$\delta \lambda$ about $\lambda_0$. Henceforth, we set $\lambda_0$ to unity.

Obviously, we are interested to see how $q$ responds to the
perturbation by the marginal operator $\cal E$. Such marginal
perturbation preserves the conformal symmetry and the value of the central
charge $c$.

{}From the definition of marginal perturbation by a marginal operator,
$\cal E$ is the field conjugate to $\delta \lambda$;
the marginal perturbation can be represented as the addition of $\delta S$
to the action, where
\beq
\delta S = \frac{\delta \lambda}{2 \pi} \int d^2 z \,
{\cal E} (z, {\overline z})
\, .
\eeq
As a result,
the variation of the correlation function of $N$ local operators
${\cal O} = \prod_i^N {\cal O}_i ( z_i, \overline{z}_i )$
is given by
\beq
\delta \langle {\cal O} \rangle
= \frac{1}{2 \pi} \int d^2 z \,
\langle {\cal E} ( z, {\overline z} ){\cal O} \rangle \, \delta \lambda
\eeq
up to first order in $\delta \lambda$.
Next, using the operator product expansion,
\beq
{\cal O}_i (z, \overline{z}) {\cal E} (w, \overline{w})
= \frac{C_i}{|z - w|^2} {\cal O}_i + \cdots \,
\eeq
and
\beq
\langle {\cal O}_i(z-w) {\cal O}_j ( \overline{z} - \overline{w} ) \rangle
= \delta_{ij}
(z - w)^{-h_i} ( \overline{z} - \overline{w} )^{-\overline{h}_i} \, ,
\eeq
we find that the conformal weights $( h_i, \overline{h}_i)$ of the
operator ${\cal O}_i ( z, \overline{z} )$ vary with respect to $\lambda$:
\beq
\delta h_i
= \delta \overline{h}_i = - C_i \delta \lambda \, ,
\eeq
Therefore, when $\cal E$ exists, we can perturb a critical point
marginally and arrive at a new critical point with a new set of
weights shifted by a finite amount $\delta \lambda$
\cite{KW,Kadanoff}. It must be emphasized that this is possible
only if $\cal E$ is an {\em integrable} marginal operator with conformal
weight $(1, 1)$.

In our case, we have
${\cal E} = \partial \varphi \overline{\partial} \varphi$, which is
a $(1,1)$ primary field. That this specific ${\cal E}$ is an integrable
marginal operator is readily verifiable. To obtain $C_i$,
the operator product expansion of the primary operator
$V_{nm}$ with $\cal E$ is of interest here.
Using the explicit expressions of the propagators
$\langle \phi (z) \phi (w) \rangle$ and $\langle \overline{\phi}
( \overline{z} ) \overline{\phi} ( \overline{w} ) \rangle$, and
applying Wick theorem, the result is
\beq
V_{nm} (z, \overline{z} ) {\cal E} ( w, \overline{w})
= - \frac{p \overline{p}}{ | z - w|^2 } V_{nm} + \cdots \, .
\eeq
Thus, the conformal weights vary with $\delta \lambda $ according to
\cite{Kadanoff-AP}
\beq
\delta h_{nm} = \delta \overline{h}_{nm} = p {\overline p} \, \delta \lambda
\, .
\label{h-nm}
\eeq

Since we have identified $\lambda$ to be the magnetic field strength,
and the conformal weight of the vertex operator is
going to yield the filling factor $\frac{1}{q}$,
the necessary condition for the formation of the plateau is
\beq
\delta q = 0 \, .
\label{d-lambda-d-q}
\eeq
The condition indicates that {\em within the interval}
$\delta \lambda$, the gradient of the straight line vanishes and that is what
a plateau should be in a plot with $q$ versus $B$.
It must be reminded that the effective theory (\ref{Gaussian})
is valid for a certain filling factor. Therefore, as long as the
system is in the FQHE phase with that filling factor,
(\ref{Gaussian}) constitutes an appropriate theory for the phenomenon.
In other words, the effective theory is good
only for a range of $\lambda$ centered at $\lambda_0$.
When it is out of the FQHE phase, (\ref{d-lambda-d-q}) need not hold any more.
Of course, as to what $\lambda_0$ is and the width of the plateau, these are
phenomenological issues and do not fall within the scope of the
effective theory. Probably $\lambda_0$ and $\delta \lambda$ depend on the
temperature, the imperfections in the heterojunction, the mobility etc.

Equations (\ref{d-lambda-d-q}) and (\ref{h-nm})
combine to give a simple result:
\beq
p = 0 \, \hspace{1cm} \hbox{\rm or} \hspace{1cm} \overline{p} = 0 \, .
\eeq
In other words, the effective field theory must be {\em chiral} (or
anti-chiral depending on the choice).
Choosing $\overline{p} = 0$, the plateau formation condition
yields a value $R_0 = \sqrt{ \frac{m}{2 n} } $ for the radius (\ref{lattice}).
and the inverse filling factor is given by $q = 2 h_{n m} = p^2$.
Clearly, the condition also spells out that $q$ must be an integer.
Thus, we see why chiral theory is called for. It is entailed by the
condition for the formation of the plateau, which is precisely
what the quantum Hall effect is all about.

However, this is not the whole story yet. The effective theory is bosonic and
its correlation functions are symmetric functions of $z_1, \, \cdots, z_N$.
It is a reflection of the fact that the ``electric" qunatum number $n$
and the ``magnetic" quantum number $m$ are both integers in (\ref{lattice}).
The consequence of having integral $n$ and $m$ is that the inverse
filling factor $q = 2 n m$
is an even number, which contradicts the Pauli principle.

The subtle aspect of a compactified bosonic theory in 2 dimensions is that
it can yield fermionic behavior as well. By now it is no surprise
(See \cite{Ginsparg}, in particular \S 8.2 for review and further references).
The essential point is that the field variable $\varphi$ describes
a target space of a torus. That means that the boundary conditions
for $\varphi$ admit spin structure. One can then revise the value of
$n$ from $\bf Z$ to $n \in {\bf Z} + \frac{1}{2}$, so that
$q$ can be an odd integer (See also \cite{MR} from the point of view of
Chern-Simons theory, which for  $U(1)$ gauge group, the level $k$ which
appears as a coupling constant of the theory is not restricted to
integers). Let $n = n' + \frac{1}{2}$ where $n'$ is an integer, and
set $m = 1$, we have $q = 2 n' + 1$ which is the inverse filling factor for
Laughlin's type FQHE. An important point of this ans\"atz is that the
chiral vertex operator $e^{ i \sqrt{q} \phi (z) }$ is single-valued under
$\phi (z) \rightarrow \phi (z) + 2 \pi R$.
Therefore, when the electron gas undergoes a transition to the FQHE phase,
$p$ in (\ref{lattice}) is no longer arbitrary; the plateau formation
condition fixes it to be an integer, and
the Pauli principle further restricts the integer to be odd.

In summary, it is now clear that the ``bosonic" Gaussian theory
can be taken as an effective theory for the FQHE of Laughlin type.
The correlation function of the {\em chiral} vertex operators with
the measure (\ref{measure}) for the Hilbert space is
identified with the Laughlin wavefunction. The discussions have shown
that the effective theory is {\em chiral} because only then will it be
consistent with the plateau formation
condition. The suggestion for taking the Gaussian model
as the effective theory is backed by the microscopic theory of
the electrons in the ``puncture" phase which involves
the Knizhnik-Zamolodchikov equation, and therefore indicative of
a $U(1)$ WZW theory (which is just the Gaussian model (\ref{Gaussian})).
The key idea has been the existence of an integrable marginal operator
for the $c=1$ conformal field theory which allows the conformal weight of
the vertex operator to vary with respect to $\lambda$.


\newpage
\begin{thebibliography}{99}

\bibitem{Fubini}
S. Fubini,
Mod. Phys. Lett. {\bf A6}, 347 (1991);
S. Fubini and C. A. L\"utken, Mod. Phys. Lett. {\bf A6}, 487, (1991).

\bibitem{MR}
G. Moore and N. Read, Nucl. Phys. {\bf B360},  362, (1991).

\bibitem{Wu}
Y. S. Wu, Phys. Rev. Lett. {\bf 52},  2103, (1984).

\bibitem{Lai-Ting}
C. H. Lai and C. Ting, Phys. Lett. {\bf B265},  341, (1991).

\bibitem{Wilczek}
For review, see {\em Fractional Statistics and Anyon Superconductivity},
ed. F. Wilczek, (World Scientific, Singapore, 1990).

\bibitem{Ting-Lai-1}
C. Ting and C. H. Lai, Mod. Phys. Lett. {\bf B5},  1293, (1991).

\bibitem{Ting-Lai-2}
C. Ting and C. H. Lai, {\it Spinning Braid Group Representation and
the Fractional Quantum Hall Effect}, NUS preprint NUS/HEP/92011,
(Jan, 1992) (hepth@xxx/~9202024).

\bibitem{KZ}
V. Knizhnik and A. Zamolodchikov, Nucl. Phys. {\bf B247},  83, (1984).

\bibitem{Laughlin}
R. B. Laughlin, Phys. Rev. Lett. {\bf 50},  1395, (1983).
See also his lecture in \cite{Wilczek}  pp. 267-268.

\bibitem{Kadanoff}
L. P. Kadanoff, J. Phys. {\bf A11}, 1399, (1978).

\bibitem{Ginsparg}
P. Ginsparg, ``Applied conformal field theory", Les Houches lectures
in {\it Fields, Strings, and Critical Phenomena}, ed. by E.
Br\'ezin and J. Zinn-Justin, North Holland (1989).

\bibitem{DVV}
R. Dijkgraaf, E. Verlinde and H. Verlinde,
Commun. Math. Phys. {\bf 115},  649, (1988).

\bibitem{Ginsparg-1}
P. Ginsparg, Nucl. Phys. {\bf B295}[FS21],  153, (1988).

\bibitem{KW}
L. P. Kadanoff and F. J. Wegner, Phys. Rev. {\bf B4},  3989, (1971).

\bibitem{Kadanoff-AP}
L. P. Kadanoff, Ann. Phys. {\bf 120},  39, (1979).

\end{thebibliography}
\end{document}